# *Computing the Image of the City*

by Bin Jiang[1]

Kevin Lynch proposed a theory of the image of the city identifying five elements that make the city legible or "imageable". The resulting mental map of the city was conventionally derived through some qualitative processes, relying on interactions with city residents to ask them to recall city elements from their minds. This paper proposes a process by which the image of the city can be quantitatively derived automatically using computer technology and geospatial databases of the city. This method is substantially based on and inspired by Christopher Alexander's living structure and Nikos Salingaros' structural order, as a city with the living structure or structural order tends to be legible and imageable. With the increasing availability of geographic information of urban environments at very fine scales or resolutions (for example, trajectories data about human activities), the proposal or solution described in this paper is particularly timely and relevant for urban studies and architectural design.

## 1. Introduction

In formulating his theory of the image of the city, Kevin Lynch (1960) suggested five city elements that constitute the mental map of the city and make the city legible or "imageable." The five city elements include paths, edges, districts, nodes, and landmarks in terms of geometric or visual ap-pearance. If represented in a cartographic map, the five elements can be put into three categories: points (nodes and landmarks), lines (paths, edges), and polygons (districts). Regardless of their appearance, city elements have one property in common—they are distinguished among hundreds, thousands, or millions of other city artifacts by their unique sizes or colors etc. Because of this common property, some researchers (Haken and Portugali 2003) have suggested a broad definition of landmarks to refer to any distinguished city elements that shape the mental map. Following Lynch's seminal work, many studies recognized that some city elements are memorable or "imageable" not because of their visual stimulus but because they possess some personal, historical, or cultural meaning (Appleyard 1969, Golledge and Spector 1978). For example, the geometric or visual appearance of a little house may be unremarkable among its surroundings, but the house is memorable because a well-known person lived there; the meaning or semantics make the little house a landmark. More personally, a location may have no particular geometric, visual, or historical significance, but could be a landmark for a specific individual because, for example, the person had a severe traffic accident there. In this paper, the mental map we refer is one shared by a majority of people.

[1] Department of Technology and Built Environment, Division of Geomatics
University of Gävle, SE-801 76 Gävle, Sweden, Email: bin.jiang@hig.se

Fundamental to the theory of the image of the city is the novel concept of legibility—a particular (visual) quality or (apparent) clarity that makes the city's layout or structure recognizable, identifiable, and eventually imageable in the human minds. Related to imagebility is the quality of a city artifact that lends a strong, vivid image. Lynch's concepts of legibility and imageability are closely related to James J. Gibson's notion of affordances developed in his direct perception theory (Gibson 1979). According to the theory, any objects in the environment afford different activities to various individuals. For example, a street affords walking or driving, a home affords living, and a park relaxing. Because of distinguished properties (geometric, topological or semantic), a city or city artifact affords remembering as it shapes a mental map in human minds (Haken and Portugali 2003).

Traditionally, a city's mental map is communicated through interviewing city residents, drawing maps, reviewing photographs, and walking in the city. The process is tedious, involving considerable collective efforts among many individuals and researchers. Recently, with computer technology, the image of the city can be studied in a quantitative manner; for example, the syntactical image of the city is based on space syntax research (Dalton and Bafna 2003) and the digital image of the city is based on three-dimensional visual fields (Eugenio and Ratti 2009). Although these studies are indeed quantitative in nature, they do not provide any solution to the derivation of the image of the city. This paper proposes that such a mental map or the image of the city can be derived automatically from the city's geospatial databases, an idea substantially based on Christopher Alexander's living structure (Alexander 2004) and Nikos Salingaros' structural order (Salingaros 2005, 2006). Both living structure and structural order characterize the intrinsic structure of the city (see Section 2). This paper uses the world city of London as a case study to illustrate the quantitative approach to the image of the city.

Section 2 of this paper briefly introduces the concepts of living structure and structural order. Section 3 presents the approach to computing the image of the city, using the mental map of London's open space as illustration. Section 4 discusses the London pattern and two ways of thinking about city, following Alexander's classic, "*A City is Not a Tree*". Finally, the conclusion includes implications and limitations of this study.

## 2. Living structure and structural order

Living structure is one of the key concepts developed by Christopher Alexander (2004) in his four-volume opus, *The Nature of Order*. To characterize the living structure, Alexander distilled 15 properties of living structure, including levels of scale, strong centers, boundaries, and simplicity among others. A living structure links to people's response consciously or subconsciously and is perceived as pleasing. Nikos Salingaros (2005) formulated a multiplicity rule to characterize the living structure. The structure or substructures form a scaling hierarchy, that is, the number of substructures (p), and the scales of the substructure (x) meet the multiplicity rule, $p*x^a$ = constant, where a is an exponent $1<a<2$. Put differently, many small scales, a few large scales, and some intermediate scales form a scaling hierarchy, and more importantly, these scales are related to conscious or subconscious human response. From an architectural design viewpoint, Salingaros (2006) postulated three laws of structural order that coincide with the 15 properties of living structure.

Underlying living structure is structural order, which is distinct from the familiar and apparent geometric order such as lines and planes, cones and triangles, circles and spheres, and squares and cubes that human beings perceive directly. In contrast, structural order, which can be expressed simply by far more small things than large ones, is neither obvious nor apparent. Structural order may appear random, arbitrary, or chaotic, but contains a hidden order underneath (c.f., Figure 1 later). This hidden order provides a new perspective for looking at city structures or forms, and a city with living structure or structural order is likely to form a vivid image in human minds.

A city is composed of many types or layers of artifacts such as streets, buildings, and parks. In the corresponding geospatial databases, these layers are often represented by different geometric primitives such as points, lines and polygons. A layer can be divided or subdivided into many units or subunits that form a hierarchical structure. For example, a layer of streets consists of many intersecting streets, and a set of buildings constitutes a coherent building complex. Inner binding forces act to form these units and subunits into a coherent and cooperative whole (Alexander 1965). Further, Nikos Salingaros (2006) stated that the different scales, from the smallest to the largest and some intermediate, should be related and work together towards a coherent whole.

A city is dead, or deadly boring, if its structure violates the multiplicity rule. Euclidean geometric thinking in pursuit of apparent geometric order is likely to produce a boring structure. A dead structure essentially lacks the hierarchy of scales or misses some levels of scale: either the smallest, the largest or the intermediate ones. Salingaros has harshly criticized modern cities or buildings that emphasize style while ignoring the underlying rules of scale that all physical, biological, and social systems tend to meet.

A city in which all buildings are the same in size or color is deadly boring (apparent order). On the other hand, a city in which every building is different from each other creates a frustrating disorder. A living city lies between the two extremes of order and disorder, and is likely to have apparent disorder but with hidden order underneath, which is where city complexity rests (Jacobs 1961). A city with apparent order or disorder is hardly able to form a mental map. Mathematically, uniform distributions characterize the two extremes. In the case of order, there is no change or variance at all, so a complete uniform distribution. In the case of disorder, the change is around an average value, leading to a Gaussian or normal distribution. For example, most of labyrinth patterns exhibit the normal distribution. The normal distribution is also called a uniform distribution in relation to some heavy-tailed distributions, which are right-skewed. The key character of the heavy-tailed distributions is that the variance or change can be categorized into to different scales that in essence form a scaling hierarchy. This characteristic is a key difference from disorder, which cannot be categorized, and eventually the change is disturbing.

The proposal outlined here for computing the image of the city is based on the assumption that real world cities bear a living structure characterized by the multiplicity rule (Salingaros 2005). The city or city artifacts are legible or imageable (eventually computable) if the city itself contains the scaling hierarchy. The following section illustrates how to compute the image of the city.

## 3. Computing the image of the city

The key to computing the image of the city is to work out the scaling hierarchy underlying the large number of city artifacts. Artifacts in the top hierarchy are likely to shape the image of the city. Thus, we adopt the head/tail division rule to derive the scaling hierarchy. The head/tail division rule states that if a variable x has values that follow a heavy-tailed distribution, then the mean of x can divide all the values into two parts: those above the mean in the head and those below the mean in the tail (Jiang and Liu 2011). Note that the head and tail are with respect to the rank-size distribution, in which the ranking from the largest to the smallest forms the x-axis, while the size represents the y-axis. This plot is the same for illustrating rank-size distributions of city size, or Zipf's law (Zipf 1949).

The computing process has several steps. First, all city artifacts are organized, layer by layer, according to artifact types, for example, streets, buildings, and parks. Current geographic information systems organize geospatial databases this same way. Second, all the city artifacts must be organized in terms of city artifacts rather than geometric primitives such as points, lines, and polygons. For example, the streets layer should not be organized as a system (or graph) of junctions or segments, but as named streets or natural streets (Jiang and Claramunt 2004, Jiang, Zhao and Yin 2008). This way of organizing is preferred because the human mind tends to perceive an entire street rather than a street segment as a city artifact. The third step is to rank the city artifacts of the same type from the largest to the smallest. Taking the street network for example, the longest street is ranked as number one, followed by the second longest, the third longest, and so on. In the corresponding rank-size plot, the distribution is heavy-tailed or skewed to far right. Because of this distribution, each street can be placed into one of two categories: those below the mean (in the tail) and those above the mean (in the head), based on the head/tail division rule (Jiang and Liu 2011). This process continues until the streets in the head no longer show a heavy-tailed distribution, a process very much like the new classification scheme, the head/tail breaks (Jiang 2012). Those streets that remain in the last head or top head are likely to form a mental map for the street network of the city. Repeat the third step for other types of city artifacts to derive additional mental maps. Finally, overlap all the resulting mental maps to complete the image of the city.

For simplicity, but without loss of generality, this computing idea can be applied to London's open space, which is mapped by more than 44,000 axial lines, each of which represents one linear space that can be perceived from a vantage point. The resulting axial map is one of the space syntax representations (Hillier and Hanson 1984). The degree of connectivity exhibits some heavy-tailed distribution (Carvalho and Penn 2004, Jiang 2009); the smallest connectivity is one, while the largest is 114. Ten levels of hierarchy are derived, as shown in Table 1 and Figure 1. This derivation process is based on the head/tail division rule described above. The reddish lines are likely to form a part of the London mental map.

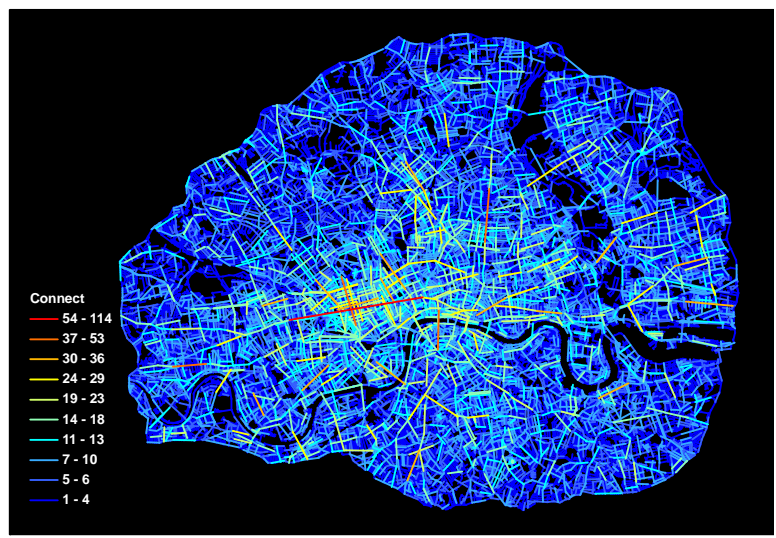

*Fig. 1 – Ten hierarchical levels for the London open space represented by the axial lines; the reddish lines constitute a part of the London mental map.*

*Tab. 1- Ten hierarchical levels and the corresponding numbers of lines (Note: I = Intervals; N = Number of lines)*

| I | 1-4 | 5-6 | 7-10 | 11-13 | 14-18 | 19-23 | 24-29 | 30-36 | 37-53 | 54-114 |
|---|-----|-----|------|-------|-------|-------|-------|-------|-------|--------|
| N | 32,357 | 6,647 | 4,007 | 845 | 444 | 102 | 48 | 16 | 4 | 1 |

The topological property connectivity was used for ranking individual lines or streets. Well-connected streets tend to be long streets; therefore, the topological property connectivity is likely to be well correlated with the geometric property length. Geometric length is also an important factor, but above all, semantic properties are the most important factors to consider. As mentioned previously, an ordinary little house is retained in a mental map not because of its prominent or distinguished geometric or topological properties, but because it has highly ranked semantic meaning. Thus, city artifacts should be ranked in the sequence of semantic, topological, and geometric, or in a combination of the three.

The pattern shown in Figure 1 provides a good illustration of structural order. The image may appear chaotic, random, or arbitrary on the surface, but a remarkable order and regularity lie underneath: there are many blue lines (the shortest lines), a few reddish ones (the longest lines), and some other colored lines (the intermediate lines). Importantly, the image evokes a sense of beauty, consciously or subconsciously linking to human response. The following section further discusses structural order, its aesthetic impacts, and related concepts.

## 4. Discussions

The pattern shown in Figure 1 can also be described as fractal. The underlying structure of the pattern meets the multiplicity rule, and exhibits the scaling hierarchy. This hierarchical pattern links to human response consciously or subconsciously. Human beings feel relaxed, comfortable, and pleasant while looking at fractal patterns. This response is evident in Richard Taylor's seminal work (2002) on bridging science and art based on the fractal properties of Jackson Pollock's poured paintings (see Figure 2). Pollock's poured paintings are remarkable fractal, providing a new way of assessing the aesthetic values of arts from a scientific point of view. The scaling hierarchy is probably the most common property in all biological, physical, and social systems. Taylor (2006) has further found that fractals help reduce physiological stress. All these insights inform the study of the image of the city. The city is essentially fractal or scaling (Batty and Longley 1994, Frankhauser 1994), and this is why the image of the city can be easily shaped in the human mind.

Built environments or cities are simply logical extension of nature, and nature—life of all kinds, landscapes, galaxies, and perhaps the entire universe—bears fundamental scaling or fractal properties across the scales from the infinite smallest to the infinite largest. This essence of nature is what Pollock captured on his canvas. The illustrated scaling hierarchy is fundamental to the living structure and the aesthetic appealing indicates the convergence of art and science. This is a new theory formulated by Christopher Alexander (2004). The increasingly available geographic information about cities provides a new means to verify Alexander's theory. Both living structure and structural order may provide scientific approaches to assessing the aesthetic values of map products. In other words, a map's quality goes beyond its surface appearance—colors, symbols, ratios—to some deep sense of order or regularity—the scaling hierarchy.

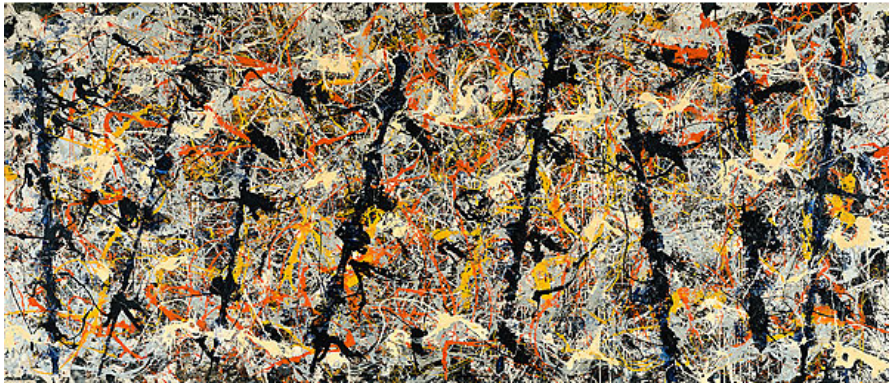

*Fig. 2 - Blue Poles: Number 11, 1952 by Jackson Pollock*

Here are two ways of thinking related to computing the image of the city: trees and networks. In his classic, "*A City is Not a Tree*", Alexander (1965) stressed that "*the city is not, cannot, and must not be a tree*". As a widely used organizational metaphor, a tree has a centralized structure, and no part of any branch is able to connect to other branches unless through their parental branches or the root. This structure would be like no member of a family being able to make friends with someone, unless through the family as a whole (Alexander 1965). On the other hand, the tree has a hierarchical structure, which, again, is a widely recognized means for conceptualizing many real world things or phenomena like formal organizations.

In contrast to the tree metaphor, a network is a decentralized structure in which every node is the center of the network, although some nodes are more centralized than others. Their statuses are not predefined as in a tree, in which there is a difference between superordinates and subordinates from the point of the view of the root, the most central node. In a network, every node can be a center, from which all other nodes can be surrounded in different levels of centrality. A complex network involving a large amount of nodes has an intrinsic scaling hierarchy, a property that trees share. As shown in Figure 1, a street network can demonstrate a scaling hierarchy. To reveal this scaling hierarchy is the main reason for using the degree of connectivity for ranking individual streets.

From the time when Lynch (1960) first proposed his theory, the derivation of the image of the city has been a qualitative conduct, relying on hu-man beings to recover city elements from their minds. Today, the derivation process can be achieved in a quantitative manner using geospatial databases of the city. This is a bold proposal, and despite the illustrations in this paper, many geographers may dislike it. Yet, as long as all the city artifacts are represented with semantic, topological, and geometric information attached, a mental map can be computed. Even without semantics, both topological and geometric information are sufficient to derive a reasonable good image of the city.

## 5. Conclusion

This paper proposes that the image of the city can be automatically computed from geospatial databases of the city. The computable image of the city lies on the fact that the city possesses a living structure. The living structure contains an intrinsic hierarchy in which there are far more small artifacts than large ones, and importantly, the intermediate scales between the smallest and largest scales work together towards a coherent whole. Thus, both legibility and imageability of city artifacts can be quantified, at least within a city, through a process of ranking (in a decreasing order) the individual city artifacts in terms of semantic, topological, and geometric information. With the ranking, we can recursively divide all city artifacts into two parts: those below the mean, and those above the mean. This division process continues until those in the head no longer demonstrate a hierarchy. Those artifacts remaining in the last head are likely to form essential parts of the mental map.

This study opens up new possibilities and opportunities for studying the image of the city in a quantitative manner, as an increasing amount of geographic information is made available through volunteered efforts (Goodchild 2007). Strictly speaking, the image of the city is that of the geospatial databases rather than of the city itself, but because the geospatial databases have become much more detailed, the mental map computed can be said to be that of the city itself. Another point worth noting is that the multiplicity rule holds mainly for facades (vertical dimension) rather than for the city plan (horizontal dimension) (Salingaros 2012). However, many previous studies on urban structure have found that the rule or scaling pattern holds remarkably true for the city plan as well. Computing the image of the city is mainly based on the horizontal information rather than on the vertical information because vertical information is rarely available in the geospatial databases. The lack of vertical information is probably one limitation of the study.

## Acknowledgement

I am grateful to Andrea De Montis for inviting me to address INPUT 2012, which spurred me to write this article, and to Itzhak Omer and Nikos Salingaros for some useful comments. This works is partially sponsored by the University of Sassari through its visiting professorship program.